\documentclass[10pt, a4paper]{article}

\usepackage{amsmath}
\usepackage{url}
\usepackage{booktabs}
\usepackage{tcolorbox}
\usepackage[table]{xcolor}
\usepackage{calc}
\newlength{\speakerwidth}
\usepackage{listings}
\usepackage[final]{lrec2026} 

\title{Transcription and Recognition of Italian Parliamentary Speeches Using Vision-Language Models}

\name{Luigi Curini$^{1}$, Alfio Ferrara$^{2}$, Giovanni Pagano$^{1}$, Sergio Picascia$^{3\ast}$}

\address{$^{1}$Università degli Studi di Milano, Department of Social and Political Sciences \\
         Via Conservatorio, 7 - 20122 Milano (Italy) \\
         luigi.curini@unimi.it, giovanni.pagano@unimi.it
         \and
         $^{2}$Università degli Studi di Milano, Department of Literary Studies, Philology and Linguistics \\
         Via Festa del Perdono, 7 - 20122 Milano (Italy) \\
         alfio.ferrara@unimi.it
         \and
         $^{3}$Università degli Studi di Milano, Department of Computer Science \\
         Via Celoria, 18 - 20133 Milano (Italy) \\
         sergio.picascia@unimi.it}

\abstract{
Parliamentary proceedings represent a rich yet challenging resource for computational analysis, particularly when preserved only as scanned historical documents. Existing efforts to transcribe Italian parliamentary speeches have relied on traditional Optical Character Recognition pipelines, resulting in transcription errors and limited semantic annotation. In this paper, we propose a pipeline based on Vision-Language Models for the automatic transcription, semantic segmentation, and entity linking of Italian parliamentary speeches. The pipeline employs a specialised OCR model to extract text while preserving reading order, followed by a large-scale Vision-Language Model that performs transcription refinement, element classification, and speaker identification by jointly reasoning over visual layout and textual content. Extracted speakers are then linked to the Chamber of Deputies knowledge base through SPARQL queries and a multi-strategy fuzzy matching procedure. Evaluation against an established benchmark demonstrates substantial improvements both in transcription quality and speaker tagging.
 \\ \newline  }
 
\begin{document}

\maketitleabstract

\section{Introduction}
\label{sec:introduction}

Parliamentary proceedings constitute one of the most valuable documentary sources for the study of political, linguistic, and social change. In the Italian context, these records chronicle nearly two centuries of transformative events. The stenographic reports produced by both chambers of the Italian Parliament, the Camera dei Deputati and the Senato della Repubblica, provide a uniquely detailed account of these developments through the verbatim transcription of political discourse.

Several initiatives have sought to make these records available in machine-readable form. Cross-national projects such as ParlaMint~\citeplanguageresource{Erjavec2022} have assembled comparable parliamentary corpora across European countries, while Italy-specific efforts, including IPSA~\citeplanguageresource{frasnelli-palmero-aprosio-2024-theres} and ItaParlCorpus~\citeplanguageresource{Cova_2025}, have produced large-scale datasets spanning extensive historical periods. However, these resources predominantly rely on traditional Optical Character Recognition (OCR) pipelines followed by rule-based heuristics for text cleaning and speaker attribution. While effective to a degree, such approaches face well-documented limitations when applied to historical documents, contributing to transcription errors and unreliable speaker identification. The latter issue is especially acute for the earlier portion of the corpus (pre-1948), where high-quality annotated data remains scarce.

Speaker identification in Italian parliamentary documents is challenging because the typographic conventions used to mark speakers vary considerably across legislatures and historical periods. Additional variability arises from the treatment of homonymous members, for whom both surname and first name are provided, and from the occasional inclusion of the speaker's institutional role alongside the surname. Taken together, these inconsistencies make rule-based speaker attribution brittle and difficult to generalise across the full historical span of the corpus.

The recent emergence of Vision-Language Models (VLMs) offers a promising alternative to pipeline-based methods. VLMs jointly process visual and textual information through unified architectures, enabling end-to-end reasoning about document layout, content, and semantics. Specialised models such as \texttt{dots.ocr}~\citep{li2025dotsocrmultilingualdocumentlayout} have demonstrated strong performance on document layout analysis and text recognition, while general-purpose models like \texttt{Qwen2.5-VL}~\citep{qwen} provide complementary capabilities in semantic understanding and contextual inference. To date, however, the potential of these models for the digitisation and annotation of historical parliamentary documents remains largely unexplored.

In this paper, we present a pipeline for the automatic transcription, semantic segmentation, and entity linking of Italian parliamentary speeches based on Vision-Language Models\footnote{The code developed for this study is openly available at \url{https://github.com/umilISLab/trips}}. Unlike previous approaches, our method leverages the visual layout of documents alongside their textual content, enabling more accurate transcription and richer semantic annotation. We evaluate the proposed pipeline against IPSA on its released benchmark, assessing both OCR quality and speaker tagging accuracy.

The remainder of this paper is organised as follows. Section~\ref{sec:related-work} reviews prior work on parliamentary corpora and vision-language models. Section~\ref{sec:methodology} describes the proposed pipeline in detail. Section~\ref{sec:evaluation} presents the experimental evaluation and discusses the results. Finally, Section~\ref{sec:conclusion} summarises the contributions and outlines directions for future work.

\section{Related Work}
\label{sec:related-work}

This section reviews prior work relevant to our contribution, organised into two main areas: parliamentary corpora with a focus on Italian resources, and vision-language models for document understanding and OCR.

\subsection{Italian Parliamentary Resources}

Parliamentary debates constitute a valuable resource for political science, linguistics, and computational social science research. The systematic collection and annotation of parliamentary proceedings has been pursued across numerous countries, resulting in large-scale corpora that enable studies of political discourse, policy preferences, and legislative behaviour. Cross-national initiatives have sought to harmonise parliamentary data across countries. The ParlaMint project~\citeplanguageresource{Erjavec2022} assembled comparable corpora from 29 European countries, containing over one billion words and covering at least the period 2015--2022, with linguistic annotations following the Universal Dependencies framework. Similarly, the ParlSpeech dataset~\citeplanguageresource{DVN/L4OAKN_2020} provides full-text corpora from various advanced democracies, though notably excluding Italy.

The digitisation and analysis of Italian parliamentary proceedings has received increasing attention in recent years. IPSA~\citeplanguageresource{frasnelli-palmero-aprosio-2024-theres} represents the most comprehensive effort to date, providing over 1.2 billion tokens of parliamentary debates from both the Camera dei Deputati and the Senato della Repubblica, spanning from 1848 to 2022. The corpus was constructed by applying Tesseract OCR to scanned documents, followed by rule-based heuristics for text cleaning and speaker tagging through fuzzy string matching against lists of parliamentarians. The ItaParlCorpus dataset~\citeplanguageresource{Cova_2025} offers a machine-readable collection of Camera dei Deputati speeches from 1948 to 2022, encompassing 470 million words with metadata including speaker identification and party affiliation. ParlaMint-It~\citeplanguageresource{Alzetta2024} contributes a manually revised treebank of Italian parliamentary debates annotated according to the Universal Dependencies framework, addressing the underrepresentation of parliamentary language varieties in syntactic resources. The IMPAQTS corpus~\citeplanguageresource{cominetti-etal-2024-impaqts} takes a multimodal approach, collecting 2.65 million tokens of political discourse from 1946 to 2023 with pragmatic annotations capturing implicit content. 

Despite these advances, existing Italian parliamentary corpora share common limitations: they predominantly rely on traditional OCR pipelines that struggle with historical document quality, employ rule-based approaches for speaker identification that cannot leverage visual layout cues, and lack fine-grained semantic annotations linking speakers to authoritative knowledge bases. Our work addresses these limitations by proposing a vision-language model pipeline that jointly performs transcription, semantic segmentation, and entity linking.

\subsection{Vision-Language Models}

Vision-Language Models (VLMs) are AI systems designed to jointly process visual and textual information~\citep{vlmsurvey}. These models typically employ a dual-encoder architecture, where separate encoders transform images and text into vector embeddings that are subsequently projected into a shared latent space. The aligned multimodal representations are then processed by a Transformer decoder, where visual embeddings serve as conditioning context for text generation. This architecture enables VLMs to perform a wide range of tasks requiring joint reasoning over images and text, including visual question answering, image captioning, and document understanding.

Optical Character Recognition has traditionally relied on pipeline approaches combining image preprocessing, layout analysis, character segmentation, and recognition~\citep{islam2017survey}. Tesseract~\citep{tesseract} remains widely used due to its extensive language support and cost-effectiveness. However, traditional OCR systems face significant challenges with historical documents, including degraded print quality, non-standard typefaces, and complex multi-column layouts. The application of VLMs to document understanding and OCR represents a paradigm shift from pipeline-based approaches to end-to-end systems capable of jointly reasoning about visual layout and textual content.

Recent benchmarks have evaluated VLMs on OCR tasks across different document types~\citep{Ouyang_2025_CVPR}. Among end-to-end models, specialised OCR-focused VLMs such as \texttt{dots.ocr}~\citep{li2025dotsocrmultilingualdocumentlayout} have achieved strong results by combining layout analysis with text recognition in a unified framework. \texttt{dots.ocr} performs document layout analysis to identify element bounding boxes and categories, followed by text transcription respecting logical reading order. General-purpose VLMs have also demonstrated competitive OCR performance when appropriately prompted. \texttt{Qwen2.5-VL-72B}~\citep{qwen} achieves results comparable to specialised systems while offering additional capabilities for semantic understanding. The combination of these complementary capabilities, accurate transcription from specialised OCR models and semantic understanding from large VLMs, enables richer annotation of parliamentary documents than previously achievable with traditional approaches.

\section{Methodology}
\label{sec:methodology}

\begin{figure}[ht!]
    \centering
    \includegraphics[width=0.9\linewidth]{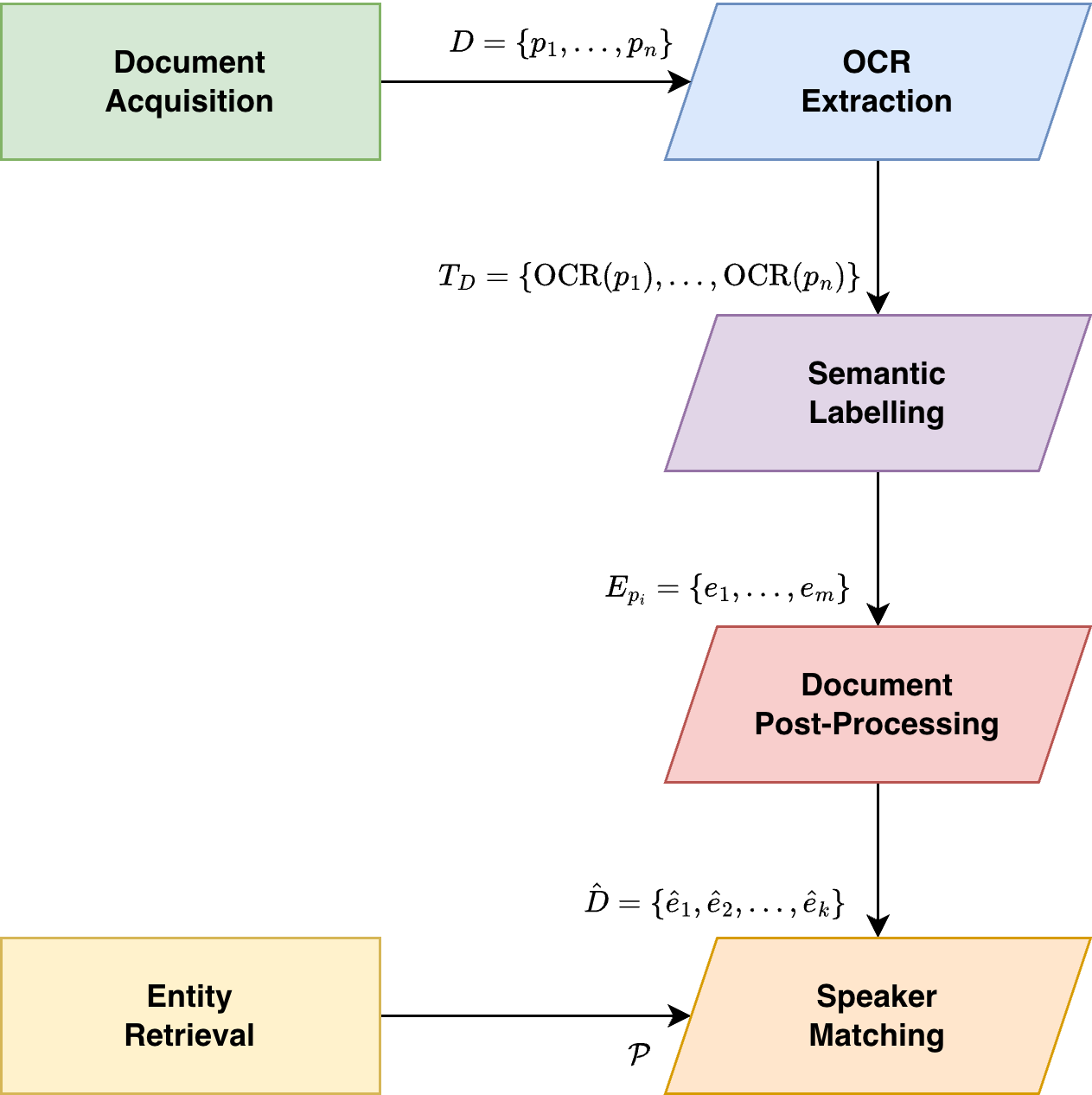}
    \caption{Pipeline diagram showing the six stages with data flow between components.}
    \label{fig:pipeline}
\end{figure}

This section presents the methodology developed for the automatic transcription and semantic labelling of Italian parliamentary session reports. The proposed pipeline transforms digitised parliamentary documents into structured, semantically annotated data, enabling downstream analyses on political discourse studies. The pipeline comprises six sequential stages: (1) Document Acquisition, (2) OCR Extraction, (3) Semantic Labelling, (4) Document Post-Processing, (5) Entity Retrieval, and (6) Speaker Matching. Figure \ref{fig:pipeline} provides a schematic overview of the complete processing pipeline.

\begin{figure}[ht!]
    \centering
    \includegraphics[width=\linewidth]{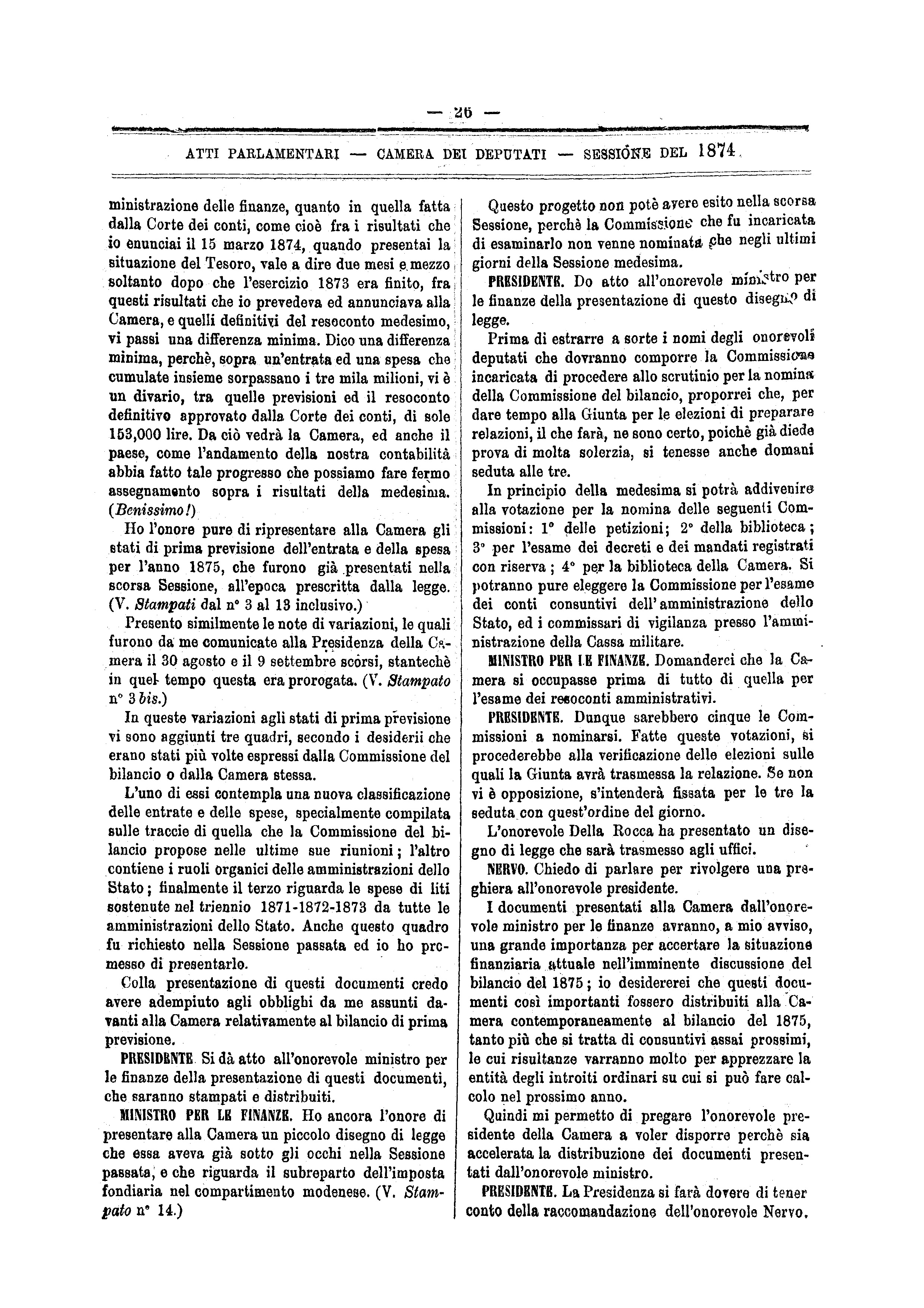}
    \caption{Excerpt from the stenographic report of the session held on November 27th 1874, Legislature 12 of the Kingdom of Italy. This page excerpt serves as the running example throughout this section.}
    \label{fig:running_example}
\end{figure}

To illustrate the transformations applied at each stage, we introduce a running example drawn from an actual parliamentary session. Figure~\ref{fig:running_example} presents an excerpt from a document page\footnote{\url{https://storia.camera.it/regno/lavori/leg12/sed004.pdf\#page=6}} that will be traced through the pipeline, demonstrating how raw visual input is progressively transformed into structured data.

\subsection{Document Acquisition}

For each legislature in the history of the Italian Parliament, we retrieved the complete list of sessions (\textit{sedute}). Each session, which can be thought of as a parliamentary meeting, is uniquely identified by a URI and associated with the specific date on which it took place. The majority of sessions are documented by a single PDF file containing the official verbatim report of the proceedings (\textit{resoconto stenografico}). These reports tend to follow a consistent structure: they contain verbatim transcriptions of parliamentary debates, with speaker names indicated at the beginning of each intervention, typically accompanied by their institutional role when applicable (e.g., {\small \textit{BIANCHI LEONARDO, MINISTER OF PUBLIC EDUCATION}}).

For the session held on November 27th 1874, belonging to Legislature 12 of the Kingdom of Italy, the following metadata was retrieved from the parliamentary portal: legislature URI ({\small http://dati.camera.it/ocd/legislatura.rdf/regno\_12}), session URI ({\small http://dati.camera.it/ocd/seduta.rdf/sr12004}), date ({\small 1874-11-27}), and document URL ({\small http://storia.camera.it/regno/lavori/leg12/sed004.pdf}).

\subsection{OCR Extraction}

The digitised parliamentary reports present several challenges for automatic text extraction. The documents exhibit a two-column layout, variable print quality depending on the historical period, and occasional multilingual content. Preliminary experiments with traditional OCR engines, such as Tesseract, revealed significant limitations when facing these issues.

To address these challenges, we employed \texttt{dots.ocr}~\citep{li2025dotsocrmultilingualdocumentlayout}, a specialised vision-language model designed for document understanding tasks. For each page $p_i$ in a document $D = \langle p_1, p_2, \ldots, p_n \rangle$, the model receives the page image and a prompt requesting document layout analysis and textual transcription in reading order. The model identifies the bounding boxes of the elements in the page, labels these elements according to the corresponding category in the page layout, and transcribes their textual content respecting the logical reading sequence.

The output of this stage undergoes a further processing to extract only the textual content, discarding layout labels. This design decision was motivated by observed inconsistencies in the labelling conventions applied by the model across pages. The semantic classification of elements was therefore delegated to the subsequent processing stage, where a more capable model could leverage both visual and textual information for consistent labelling.

Formally, let $\text{OCR}(p_i)$ denote the function mapping a page image to its preliminary textual transcription. The output of this stage for a document $D$ is the sequence of page-level transcriptions $T_D = \langle \text{OCR}(p_1), \text{OCR}(p_2), \ldots, \text{OCR}(p_n) \rangle$.

Processing the documents with a specialised model allows us to achieve high transcription quality and successfully preserves reading order, as demonstrated in the following example:

\begin{lstlisting}
- 26 -

# ATTI PARLAMENTARI - CAMERA DEI DEPUTATI - SESSIONE DEL 1874.

ministrazione delle finanze, quanto in quella fatta dalla Corte dei conti, come cioe' fra i risultati che io enunciai il 15 marzo 1874, quando presentai la situazione del Tesoro, vale a dire due mesi e mezzo soltanto dopo che l'esercizio 1873 era finito, fra questi risultati che io prevedeva ed annunciava alla Camera, e quelli definitivi del resoconto medesimo, vi passi una differenza minima.
\end{lstlisting}

\subsection{Semantic Labelling}

This stage involves the employment of \texttt{Qwen-VL2.5-72B}, a large-scale vision-language model, to perform transcription refinement, textual segmentation, element type classification, and speaker identification. For each page $p_i$, the model receives three inputs: (i) the page image, (ii) a structured prompt that specifies the annotation schema and task requirements, and (iii) the preliminary OCR transcription $\text{OCR}(p_i)$.

The prompt instructs the model to segment the page content into discrete elements, representing a segmented textual unit, each characterised by three attributes:
\begin{itemize}
    \item \textit{type}: a categorical label that indicates the role of the element on the page. The type is drawn from the set of labels: page-header, section-header, text, note, footnote, table. Elements labelled as text represent the main content of the report, i.e. the speeches, and are those that are typically assigned to speakers. The note type designates parenthetical remarks within the main text that record non-verbal events or reactions during the session (e.g., \textit{Hilarity}, \textit{The Chamber approves});
    \item \textit{content}: the verbatim transcription of the element, preserving original punctuation and orthography. Line-broken sentences within a single element are asked to be reconstructed as continuous text;
    \item \textit{speaker}: the name of the person speaking, including their institutional role if explicitly mentioned. For elements that do not constitute speeches (headers, footnotes, notes, tables) or that report neutral content such as article text being read aloud, this field is set to "none". When a speech continues from a previous element on the same page without an explicit speaker indication, the model is instructed to infer the speaker from context. If inference is not possible (e.g., text continuing from a previous page), the field is marked as "unknown".
\end{itemize}

To guide the model's behaviour on challenging cases, the prompt includes a synthetic example, which aggregates multiple difficult scenarios: text continuation from previous pages, speaker role annotations, parenthetical notes interrupting speeches, footnote references, and section transitions. This one-shot example demonstrates the expected output format and handling of edge cases. The model is instructed to return its output in JSON format, producing for each page a sequence of annotated elements, $E_{p_i} = \langle e_1, e_2, \ldots, e_m\rangle$, where each element $e_j = (\text{type}_j, \text{content}_j, \text{speaker}_j)$.

The following example shows the beginning of a structured output produced by the semantic labelling stage:

\begin{lstlisting}
    {
        "speaker": "none",
        "type": "page-header",
        "content": "- 26 -"
    },
    {
        "speaker": "none",
        "type": "page-header",
        "content": "ATTI PARLAMENTARI - CAMERA DEI DEPUTATI - SESSIONE DEL 1874."
    },
    {
        "speaker": "unknown",
        "type": "text",
        "content": "ministrazione delle finanze, quanto in quella fatta dalla Corte dei conti, come cioe' fra i risultati che io enunciai il 15 marzo 1874, quando presentai la situazione del Tesoro, vale a dire due mesi e mezzo soltanto dopo che l'esercizio 1873 era finito, fra questi risultati che io prevedeva ed annunciava alla Camera, e quelli definitivi del resoconto medesimo, vi passi una differenza minima."
    }
\end{lstlisting}

\noindent In this example, the third element is a speech fragment whose speaker is marked as \texttt{"unknown"}. The text begins without any speaker heading, indicating that the speech started on a preceding page. Because the model processes each page independently, it has no access to the previous page's context and therefore cannot determine the identity of the speaker. Such cases are resolved in the subsequent phase.

\subsection{Document Post-Processing}

The outputs from the vision-language model undergo a post-processing phase to address artefacts arising from page-level processing and to resolve cross-page dependencies. This stage performs the following operations.

\textbf{Hyphenation Resolution}. Words hyphenated at line or column breaks are rejoined by detecting and removing mid-word hyphens followed by whitespace patterns indicative of line continuation.

\textbf{Cross-Page Element Merging}. Elements truncated at page boundaries are identified and merged with their continuations on subsequent pages. This is achieved by detecting incomplete sentences (lacking terminal punctuation) at page endings and matching them with elements marked as continuations (speaker "unknown" with type "text") at the beginning of subsequent pages.

\textbf{Role Extraction}. Institutional roles embedded within speaker names or element content are extracted and normalised. When a speaker is identified with an accompanying role, the role is parsed and separated by a comma from the speaker name.

\textbf{Speaker Continuity Inference}. For elements where the speaker could not be determined during page-level processing (marked as "unknown"), we implement a backward-looking inference mechanism. The algorithm traverses the document in reading order, propagating speaker attribution from the most recent explicitly identified speaker until a discontinuity marker is encountered. A discontinuity marker is represented by the appearance of a new section header, indicating a thematic break in proceedings. This ensures that speeches spanning multiple pages are correctly attributed to their speakers while respecting the logical structure of parliamentary proceedings.

The output of this stage is a document-level sequence of processed elements: $\hat{D} = \langle \hat{e}_1, \hat{e}_2, \ldots, \hat{e}_k\rangle$.

\begin{figure}[ht!]
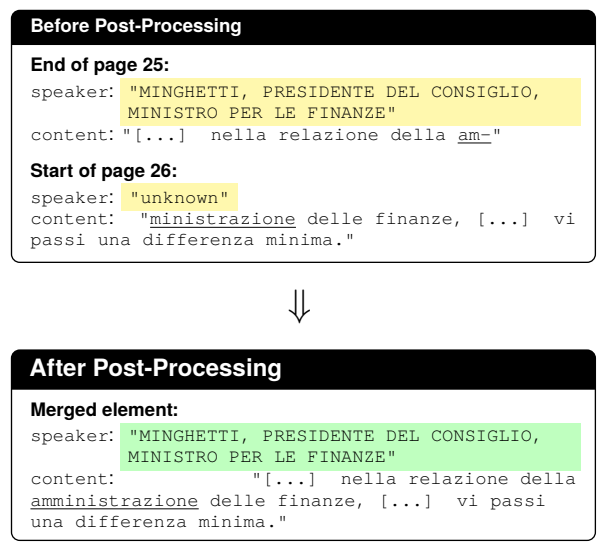

\centering
\begin{tcolorbox}[colback=white, colframe=black, title={\textbf{Before Post-Processing}}, fonttitle=\scriptsize, boxrule=0.5pt, left=4pt, right=4pt, top=2pt, bottom=2pt]
\scriptsize
\textbf{End of page 25:}\\
\texttt{speaker}: \colorbox{yellow!40}{\parbox[t]{\dimexpr\linewidth-2\fboxsep-\speakerwidth\relax}{\texttt{"MINGHETTI, PRESIDENTE DEL CONSIGLIO, MINISTRO PER LE FINANZE"}}}\\
\texttt{content}: \texttt{"[...] nella relazione della \underline{am-}"}

\medskip
\textbf{Start of page 26:}\\
\texttt{speaker}: \colorbox{yellow!40}{\texttt{"unknown"}}\\
\texttt{content}: \texttt{"\underline{ministrazione} delle finanze, [...] vi passi una differenza minima."}
\end{tcolorbox}

\smallskip
{\Large$\Downarrow$}
\smallskip

\begin{tcolorbox}[colback=white, colframe=black, title={\textbf{After Post-Processing}}, fonttitle=\small, boxrule=0.5pt, left=4pt, right=4pt, top=2pt, bottom=2pt]
\scriptsize
\textbf{Merged element:}\\
\texttt{speaker}: \colorbox{green!25}{\parbox[t]{\dimexpr\linewidth-2\fboxsep-\speakerwidth\relax}{\texttt{"MINGHETTI, PRESIDENTE DEL CONSIGLIO, MINISTRO PER LE FINANZE"}}}\\
\texttt{content}: \texttt{"[...] nella relazione della \underline{amministrazione} delle finanze, [...] vi passi una differenza minima."}
\end{tcolorbox}
\caption{Effect of post-processing on a cross-page speech fragment. Highlighted regions show speaker inference (\colorbox{yellow!40}{unknown} $\rightarrow$ \colorbox{green!25}{resolved}); underlined text shows hyphenation resolution.}
\label{fig:postprocessing}
\end{figure}

Figure~\ref{fig:postprocessing} illustrates the effect of post-processing on the running example. The speech fragment on page~26, originally marked with an \texttt{"unknown"} speaker, is merged with the incomplete element at the end of page~25: the hyphenated word is rejoined, the content is concatenated, and the speaker identity is propagated from the preceding context.

\subsection{Entity Retrieval}

To enable linking of extracted speakers to authoritative records, we retrieve from the Italian Chamber of Deputies knowledge base all individuals who held parliamentary positions on the date of each session. The knowledge base, accessible via a SPARQL endpoint\footnote{\url{https://dati.camera.it/sparql}}, contains structured information about deputies, government members, parliamentary officers, and members of institutional organs. For a session occurring on date $d$, we execute a series of SPARQL queries to retrieve the set of active entities $\mathcal{P}_d = \mathcal{P}_d^{\text{dep}} \cup \mathcal{P}_d^{\text{gov}} \cup \mathcal{P}_d^{\text{org}} \cup \mathcal{P}_d^{\text{off}}$, where $\mathcal{P}_d^{\text{dep}}$ denotes deputies, $\mathcal{P}_d^{\text{gov}}$ government members, $\mathcal{P}_d^{\text{org}}$ members of parliamentary organs, and $\mathcal{P}_d^{\text{off}}$ parliamentary officers active on date $d$.

For each entity $p \in \mathcal{P}_d$, we retrieve: the unique URI serving as the canonical identifier, the full name, surname, given name, and the set of institutional roles held on the relevant date. This knowledge base is linked to Wikidata, enabling subsequent enrichment with additional biographical and political information such as party affiliation and demographic attributes.

On November 27th 1874, we have 557 active parliamentarians. For each entity, we retrieve the URI, name, and roles:

\begin{lstlisting}
{
    "uri": "http://dati.camera.it/ocd/persona.rdf/pr3028",
    "fullname": "MARCO MINGHETTI",
    "name": "MARCO",
    "surname": "MINGHETTI",
    "dep": true,
    "gov": [
        "MINISTRO: MINISTERO DELLE FINANZE", 
        "PRESIDENTE: PRESIDENZA DEL CONSIGLIO"
    ],
    "org": [],
    "off": []
}
\end{lstlisting}

\subsection{Speaker Matching}

The final stage establishes correspondences between extracted speaker names and knowledge base entities. Given the historical nature of the corpus and the variety of conventions used to identify speakers in parliamentary proceedings, a simple string matching approach proves insufficient. Speaker names may appear as surnames only, as full names, with abbreviated given names, or solely by institutional role (e.g., \textit{PRESIDENT}, \textit{MINISTER OF THE INTERIOR}). Furthermore, multiple individuals may share the same surname within a single legislature. We therefore implement a multi-strategy matching pipeline that applies increasingly sophisticated matching criteria. The pipeline processes each unique speaker name extracted from a document and attempts to link it to entities in $\mathcal{P}_d$.

\textbf{Generic Speaker Filtering}. Certain speaker designations refer to collective or anonymous voices (e.g., \textit{VOICES}, \textit{A DEPUTY}) and are excluded from entity linking by pattern matching.

\textbf{Role-Based Identification}. When a speaker is identified solely by institutional role, the system queries the entity set for individuals holding that role on the session date. A special case handles the session president (\textit{PRESIDENT}): the document's first page typically contains a header followed by the presiding officer's name, which is extracted and used to disambiguate among members of the \textit{Presiding Committee}.

\textbf{Name Matching with Fuzzy Strategies}. For speakers identified by name, the system applies a cascade of fuzzy matching algorithms with decreasing stringency. The matching process employs multiple similarity metrics, including token-based ratios, partial matching, and token set comparisons, applied iteratively to both surnames and full names. Candidate entities are those exceeding a configurable similarity threshold.

\textbf{Disambiguation}. When multiple candidate entities remain after initial matching, a disambiguation cascade is applied:

\begin{enumerate}
    \item score-based ranking: candidates are ranked by their fuzzy matching scores; if a single candidate achieves the highest score, it is selected;
    \item role matching: if the speaker's role was extracted, candidates whose roles match the extracted role are prioritised;
    \item full name similarity: remaining ties are broken by computing similarity between the candidate's full name and the extracted speaker name;
    \item abbreviated name handling: for speaker names containing initials (e.g., "G. ROSSI"), the system generates abbreviated forms of candidate names and compares them;
    \item contextual mention: candidates whose full names appear elsewhere in the document text are favoured;
    \item weighted edit distance: a weighted Levenshtein distance is computed, assigning lower substitution costs to vowel-vowel substitutions to account for spelling variations common in historical documents.
\end{enumerate}

If disambiguation succeeds, the speaker is linked to the entity's URI; otherwise, all high-scoring candidates are retained as potential matches. In a second pass, unresolved speakers are compared against successfully resolved speakers from the same document. If an unresolved speaker name exhibits high similarity to a resolved one, the linking from the resolved speaker is propagated.

The output of the complete pipeline is a JSON file for each session document, containing the sequence of annotated elements with speaker entity URIs where linking succeeded. This structured representation enables direct integration with the parliamentary knowledge base and, through its linkage to Wikidata, facilitates enrichment with political party affiliations, biographical data, and cross-references to external resources for comprehensive political discourse analysis.

\begin{lstlisting}
    {
        "speaker": "MINGHETTI, PRESIDENTE DEL CONSIGLIO, MINISTRO PER LE FINANZE",
        "type": "text",
        "content": "In questa occasione credo che la Camera sara' contenta di sentire quello che gia' vedra' distintamente tanto nella relazione della amministrazione delle finanze, quanto in quella fatta dalla Corte dei conti, come cioe' fra i risultati che io enunciai il 15 marzo 1874, quando presentai la situazione del Tesoro, vale a dire due mesi e mezzo soltanto dopo che l'esercizio 1873 era finito, fra questi risultati che io prevedeva ed annunciava alla Camera, e quelli definitivi del resoconto medesimo, vi passi una differenza minima.",
        "speaker_uri": "http://dati.camera.it/ocd/persona.rdf/pr3028",
        "wikidata_uri": "Q597155"
    }
\end{lstlisting}

\section{Evaluation}
\label{sec:evaluation}

To assess the effectiveness of the proposed pipeline, we conduct a comparative evaluation against IPSA~\citeplanguageresource{frasnelli-palmero-aprosio-2024-theres}, a previously published system for Italian parliamentary corpus construction. We evaluate both the OCR transcription quality and the speaker tagging accuracy using the benchmark dataset released by the authors.

\subsection{Evaluation Setup}

\textbf{Benchmark Dataset}. The evaluation relies on the benchmark dataset released alongside IPSA. The dataset consists of 60 scanned parliamentary pages paired with manual transcriptions, which serve as the ground truth for OCR evaluation. These pages span the period from 1848 to 1996 and cover documents from both the Camera dei Deputati and the Senato della Repubblica across multiple legislatures. For the speaker tagging task, annotations are available for 58 of these 60 pages. We retrieved the page images and reference annotations from the authors' repository and processed the same page images through our pipeline.

\textbf{Metrics}. We adopt Word Error Rate (WER) and Character Error Rate (CER) for OCR quality assessment. Both metrics quantify the edit distance between the predicted transcription and the ground truth:
\begin{equation}
    \text{WER} = \frac{S + D + I}{N}, \quad \text{CER} = \frac{S + D + I}{N}
\end{equation}
where $S$ denotes the number of substitutions, $D$ the number of deletions, $I$ the number of insertions, and $N$ the total number of words (for WER) or characters (for CER) in the ground truth. Lower values indicate higher transcription quality.

For the tagging task, we report Precision, Recall, and F1 score. A true positive corresponds to a correctly identified speaker-entity link, a false positive to an incorrect identification, and a false negative to a missed identification.

\textbf{Experimental Configuration}. We employed \texttt{dots.ocr}\footnote{\url{https://github.com/rednote-hilab/dots.ocr}} for the initial OCR processing, and then \texttt{Qwen-VL2.5-72B}\footnote{\url{https://huggingface.co/RedHatAI/Qwen2.5-VL-72B-Instruct-FP8-dynamic}} with 8-bit quantisation to enable semantic inference. The experiments were run on a Nvidia H100 NVL 94 GB graphics card.

\subsection{OCR Evaluation}

Table~\ref{tab:ocr_results} presents the OCR evaluation results on the 60 benchmark pages, comparing the Tesseract baseline used in IPSA against the two stages of our pipeline: (i) \texttt{dots.ocr} alone, and (ii) \texttt{dots.ocr} followed by \texttt{Qwen-VL} transcription refinement.

\begin{table}[ht!]
    \centering
    \begin{tabular*}{1\linewidth}{@{\extracolsep{\fill}}lcc}
    \toprule
    \textbf{Method} & \textbf{CER} & \textbf{WER} \\
    \midrule
    Tesseract (IPSA) & 0.030 & 0.071 \\
    dots.ocr & 0.031 & 0.050 \\
    dots.ocr + Qwen-VL & \textbf{0.009} & \textbf{0.024} \\
    \bottomrule
    \end{tabular*}
    \caption{OCR evaluation results on the 60-page benchmark. Lower values indicate better performance.}
    \label{tab:ocr_results}
\end{table}

The results indicate that when applied in isolation, \texttt{dots.ocr} achieves a comparable CER to Tesseract (0.031 vs.\ 0.030) while yielding a notable improvement at the word level, reducing WER by approximately 30\% (from 0.071 to 0.050). The integration of \texttt{Qwen-VL} for transcription refinement delivers substantial gains: the complete pipeline achieves an error reductions of approximately 70\% in CER and 66\% in WER compared to the Tesseract baseline.

\subsection{Tagging Evaluation}

Table~\ref{tab:tagging_results} presents the speaker tagging evaluation results, comparing the rule-based approach of IPSA against our pipeline. 

The released benchmark provides only page images without the relative session metadata. Since our pipeline requires this information, in particular the session date for Senate document, in order to retrieve the set of active parliamentarians, we restrict our evaluation to the 29 \textit{Chamber of Deputies} pages for which the date could be reliably extracted from the filename.

\begin{table}[ht!]
\centering
\begin{tabular*}{1\linewidth}{@{\extracolsep{\fill}}lccc}
\toprule
\textbf{Method} & \textbf{Precision} & \textbf{Recall} & \textbf{F1} \\
\midrule
IPSA (Global) & \textbf{0.939} & 0.880 & 0.909 \\
Ours (Global) & 0.885 & \textbf{0.970} & \textbf{0.925} \\
\midrule
IPSA (Pre-WW2) & \textbf{0.953} & 0.850 & 0.898 \\
Ours (Pre-WW2) & 0.883 & \textbf{0.970} & \textbf{0.924} \\
\midrule
IPSA (Post-WW2) & \textbf{0.916} & 0.942 & 0.929 \\
Ours (Post-WW2) & 0.892 & \textbf{0.971} & \textbf{0.930} \\
\bottomrule
\end{tabular*}
\caption{Speaker tagging evaluation results.}
\label{tab:tagging_results}
\end{table}

Our pipeline achieves a higher global F1 score, driven by substantially higher Recall, while the IPSA baseline retains an advantage in Precision. The lower precision of our method is partly explained by a systematic difference in segmentation granularity: when a speech is interrupted by a parenthetical note (e.g., \textit{The Chamber approves}), our pipeline splits the surrounding text into two distinct speech elements, whereas the ground truth treats it as a single continuous speech. This produces a higher number of predicted speech segments and, consequently, additional false positives.

A notable finding is that our pipeline exhibits no significant performance gap between the Pre-WW2 and Post-WW2 subsets, whereas the IPSA baseline shows a marked sensitivity to document quality across historical periods. This degradation is consistent with the lower typographic quality of Pre-WW2 documents, which exhibit greater variability in print clarity, adversely affecting both OCR accuracy and the reliability of the downstream tagging task. The stability in performance of our VLM-based approach suggests that the superior transcription quality does not merely benefit the exploitation of the text itself, but also carries over to subsequent tasks, keeping them uneffected by the degradation inherent in older source documents.

\subsection{Limitations}

Several factors limit the direct comparability of tagging performance between our pipeline and the IPSA baseline and may partially account for observed differences. As noted above, the absence of session date metadata in \textit{Senate} document filenames restricted the tagging evaluation to the \textit{Chamber of Deputies} subset. For these pages, the date was extracted from the filename and used to query the set of active parliamentarians. In a full-corpus processing scenario, session dates would be readily available from the document metadata, removing this constraint entirely.

Furthermore, the fragmented nature of the benchmark, consisting of isolated pages rather than complete documents, prevents our pipeline from demonstrating its full capabilities in cross-page speaker inference. In a standard workflow, a speech that begins on a previous page and continues onto the current one would be attributed to the correct speaker through the Speaker Continuity Inference mechanism. In the benchmark setting, however, no previous page is available, and the ground truth accordingly leaves the first speech unlabelled when no explicit speaker heading appears on the page. Our system is therefore evaluated without exercising one of its core design strengths.

A related limitation concerns the identification of the presiding officer. In a complete document, the President of the session is explicitly named on the first page, allowing our system to propagate this identity throughout subsequent pages. Because benchmark pages are processed in isolation without access to this introductory context, our method must approximate the presiding officer based solely on the legislature, a heuristic that is prone to error when mid-term changes in presidency occur or when a Vice President substitutes for the main President.

Finally, while our approach requires significantly higher computational resources, necessitating GPUs for Vision-Language Model inference, we argue that this cost is justified by the superior quality of the final output, which is crucial for enabling accurate downstream NLP tasks such as topic modelling or sentiment analysis.

\section{Conclusion}
\label{sec:conclusion}

In this paper, we presented a pipeline for automatic transcription, semantic segmentation, and entity linking of Italian parliamentary speeches based on Vision-Language Models. The proposed approach combines a specialised OCR model (\texttt{dots.ocr}) with a large-scale VLM (\texttt{Qwen2.5-VL-72B}) to jointly perform text extraction, element classification, and speaker identification, followed by entity retrieval from the Chamber of Deputies knowledge base and a multi-strategy fuzzy matching procedure for linking speakers to records.

Evaluation against the IPSA benchmark demonstrated that the VLM-based pipeline achieves substantial improvements in transcription quality, with relative error reductions of approximately 70\% in Character Error Rate and 66\% in Word Error Rate compared to the Tesseract baseline. The results of the speaker tagging showed competitive performance, with our method achieving higher F1 scores than the baseline across all temporal subsets. Notably, the pipeline exhibited consistent performance in both pre- and post-WW2 documents, suggesting robustness to the considerable variation in source document quality across historical periods. The evaluation on isolated benchmark pages represents a conservative estimate of performance, as the pipeline's cross-page inference mechanisms could not be fully exploited in this setting.

We are currently preparing the public release of a dataset spanning from 1848 to 1948 extracted with the proposed approach. The released data will include OCR transcriptions, structural annotations, and speaker-entity links in a structured, machine-readable format. Processing complete documents rather than isolated pages improves speaker tagging accuracy, as the system can fully exploit cross-page inference mechanisms such as session president identification and speaker continuity propagation. The link to external knowledge bases, such as Wikidata, opens the possibility of enriching the corpus with structured metadata, including party affiliations and demographic attributes, enabling richer downstream analyses in political science and computational social science.

\section*{Acknowledgements}
Computational resources provided by INDACO Core facility, which is a project of High Performance Computing at the University of MILAN \url{http://www.unimi.it}

\section*{Bibliographical References}
\label{sec:reference}

\bibliographystyle{lrec2026-natbib}
\bibliography{refs}

\section*{Language Resource References}
\label{lr:ref}
\bibliographystylelanguageresource{lrec2026-natbib}
\bibliographylanguageresource{refs-lr}

\end{document}